\documentclass[11pt]{amsart}
\usepackage{geometry}                % See geometry.pdf to learn the layout options. There are lots.
\usepackage{graphicx}
\usepackage{stmaryrd}
\usepackage{amssymb}
\usepackage{amsmath, cancel, centernot }
\usepackage[mathscr]{eucal}
\usepackage{mathtools}
\usepackage{epstopdf}
\title[Invariant Set Theory]{Invariant Set Theory}
\author{T.N.Palmer\\ Department of Physics, University of Oxford, UK}
\email{tim.palmer@physics.ox.ac.uk}                                     
\makeatletter
\newcommand\be{\@ifstar{\[}{\begin{equation}}}
\newcommand\ee{\@ifstar{\]}{\end{equation}}}
\newcommand\bp{\begin{pmatrix}}
\newcommand\ep{\end{pmatrix}}

\makeatother
\begin{document}
\bibliographystyle{plain}
%\maketitle

\begin{abstract}
Invariant Set Theory (IST) is a realistic, locally causal theory of fundamental physics which assumes a much stronger synergy between cosmology and quantum physics than exists in contemporary theory. In IST the (quasi-cyclic) universe $U$ is treated as a deterministic dynamical system evolving precisely on a measure-zero fractal invariant subset $I_U$ of its state space. In this approach, the geometry of $I_U$, and not a set of differential evolution equations in space-time $\mathcal M_U$, provides the most primitive description of the laws of physics. As such, IST is non-classical. The geometry of $I_U$ is based on Cantor sets of space-time trajectories in state space, homeomorphic to the algebraic set of $p$-adic integers, for large but finite $p$. In IST, the non-commutativity of position and momentum observables arises from number theory - in particular the non-commensurateness of $\phi$ and $\cos \phi$. The complex Hilbert Space and the relativistic Dirac Equation respectively are shown to describe $I_U$, and evolution on $I_U$, in the singular limit of IST at $p=\infty$; particle properties such as de Broglie relationships arise from the helical geometry of trajectories on $I_U$ in the neighbourhood of $\mathcal M_U$. With the p-adic metric as a fundamental measure of distance on $I_U$, certain key perturbations which seem conspiratorially small relative to the more traditional Euclidean metric, take points away from $I_U$ and are therefore unphysically large. This allows (the $\psi$-epistemic) IST to evade the Bell and Pusey \emph{et al} theorems without fine tuning or other objections.  In IST, the problem of quantum gravity becomes one of combining the pseudo-Riemannian metric of $\mathcal M_U$ with the p-adic metric of $I_U$. A generalisation of the field equations of general relativity which can achieve this is proposed, leading to new suggestions the nature of the dark universe, space-time singularities, and the fate of information in black holes. Other potentially testable consequences are discussed.
\end{abstract}

\maketitle

\section{Introduction}

Ongoing debates about the nature of both the dark and inflationary universe, and of the fate of information in black holes, are testament to the fact that we have yet to synthesise convincingly our theories of quantum and gravitational physics. Whilst such synthesis will likely require some revision to both quantum theory and general relativity, the basic premise of this paper is that by far the more radical revision will have to be to quantum theory, its success on laboratory scales notwithstanding. Here we propose a realistic locally causal theory of quantum physics, Invariant Set Theory, from which quantum theory arises as a singular limit \cite{Berry}; it is $\psi$-epistemic in the sense that the variable of the theory which corresponds to the quantum wavefunction, merely describes some distribution over deeper ontic states. The theory differs from all previous attempts in this direction in that it is assumed that there is a much greater synergy between cosmology and quantum physics than is generally assumed in traditional approaches to formulating laws of physics. That is to say, it is assumed that the nature of the world around us is as much determined by `top-down' constraints \cite{Ellis} from cosmology, as from reductionist `bottom-up' constraints from elementary-particle physics. 

The background to Invariant Set Theory is the seminal work of Lorenz \cite{Lorenz:1963} who showed that the $t \rightarrow \infty$ asymptotic behaviour of a (classical) nonlinear dynamical system is characterised by a measure-zero fractal geometry in the system's state space. As it turned out, such fractal attractors characterise a generic class of nonlinear dynamical systems \cite{Strogatz}. With this in mind, we imagine the (mono-)universe $U$ to be described by a quasi-cyclic cosmology, evolving repeatedly over multiple aeons \cite{SteinhardtTurok, Penrose:2010}, such that the space-time trajectory of $U$ (over all aeons) describes a measure-zero fractal subset $I_U$ in the state-space of $U$. The conditions for such a concept to be dynamically consistent are discussed below (see especially Section \ref{gravity}). This `Cosmological Invariant Set Postulate' \cite{Palmer:2009a, Palmer:2014} implies that the geometry of $I_U$ should be considered a more primitive expression of the laws of physics than some set of differential evolution equations for fields within space-time $\mathcal M_U$, the latter characterising the traditional reductionist approach to formulating fundamental physical theory. The relationship between these two perspectives is mathematically non-trivial: indeed in nonlinear dynamical systems theory, the relationship between differential evolution equations and fractal invariant set geometry is non-computational \cite{Blum, Dube:1993}. 

Such a \emph{volte face} has some immediate non-classical consequences. For example, any expression of the laws of physics as deterministic differential equations for fields in $\mathcal M_U$ must necessarily be incomplete: by hypothesis, the behaviour of these fields is influenced by the geometry $I_U$ in the state-space neighbourhood of $\mathcal M_U$ and not just by fields within $\mathcal M_U$ itself. In the context of quantum physics, it will be shown in Sections \ref{elements} and \ref{applications} that such geometric influence is the basis for the Heisenberg Uncertainty Principle, the de-Broglie relations and of particle-antiparticle pairing. In the context of gravitational physics, it is postulated in Section \ref{gravity} that such geometric influence may be the basis of the dark universe (both dark matter and dark energy). 

Another consequence is that if the ontic states of the universe are precisely those that lie on $I_U$, any state-space perturbation which takes such a state off $I_U$ is, by definition, inconsistent with the laws of physics. As discussed below, such perturbations are required to show that a putative realistic locally causal theory is constrained by some quantum no-go theorem (e.g. \cite{KochenSpecker, Bell, CHSH, Pusey}). By contrast, conventional `hidden-variable' approaches (deterministic or stochastic), where the space of ontic states is a continuum set, will necessarily be constrained by these theorems. The Bell \cite{Bell}  and Pusey \emph{et al} \cite{Pusey} theorems are the focus of attention below. 

As with Riemannian geometry, a key quantity needed to describe the geometry of $I_U$ is its metric. Traditionally, it has been assumed that the relevant state-space metric is Euclidean just as in space-time. Here we show that, because the geometry of $I_U$ is fractal, the appropriate metric between two space-times on $I_U$ is instead what is referred to as the `locally p-adic' metric $g_p(\mathcal M_U, \mathcal M_U')$  \cite{Robert}, where $p$ is a large (Fermat) prime.  By Ostrowski's theorem the Euclidean and p-adic metrics are the only inequivalent metrics in number theory. The reason for the relevance of the p-adic metric, as discussed in Section \ref{elements}, is the homeomorphism between the algebraic set of p-adic integers and the types of Cantor set $C(p)$ which underpin $I_U$. In particular, for large $p$, two space-times which are close in the Euclidean metric, but where one does not lie on $I_U$, are far apart in the p-adic metric. Importantly, it is shown below that Invariant Set Theory is not fine-tuned with respect to small p-adic amplitude perturbations, even though it may (erroneously) seem conspiratorially fine tuned with respect to the Euclidean metric. Because of this Invariant Set Theory can negate the Bell and Pusey et al no-go theorems without violating local causality or realism. 

It is shown that evolution on $I_U$ can be readily described by the relativistic Dirac equation (in the singular limit $p=\infty$). The problem of unifying quantum and gravitational physics then becomes one of synthesising the pseudo-Riemannian metric of $\mathcal M_U$ and the p-adic metric of $I_U$ into a single geometric theory. A possible route for such a synthesis, based on a comparatively minor revision to general relativity theory is outlined in Section \ref{gravity}. 

In Section \ref{elements}, the elements of Invariant Set Theory are described and developed. In Section \ref{applications}, the theory is applied to some of the standard quantum phenomena, focussing on quantum no-go theorems and quantum interference. In Section \ref{gravity}, some speculations are made about the application of Invariant Set Theory to situations where gravity is important. To some extent, these Section can be read independently of one another. Hence the results discussed in Section \ref{applications} could provide motivation for a deeper study of Section \ref{elements}. Some conclusions are given in Section \ref{conclusions}. 

\section{Elements of Invariant Set Theory}
\label{elements}
\subsection{The Cantor Set $C(p)$ and Basins of Attraction on $I_U$}
\label{prelim}

As mentioned in the Introduction, a quasi-cyclic cosmology is assumed, where $U$ evolves through multiple aeons \cite{SteinhardtTurok, Penrose:2010}. If $I_U$ is literally fractal, then these aeons never repeat. On the other hand, there is nothing which fundamentally prevents $I_U$ from being some `fat fractal' where evolution does repeat after a very large but finite number of aeons, and $I_U$ is in fact a limit cycle. However, the mathematics is more straightforward if a precise fractal is assumed (in a very similar way to where the mathematics underpinning classical dynamics is more straightforward if it assumed that phase space is some Euclidean space $\mathbb{R}^{2n}$ than some discretisation thereof). 

The geometry of  $I_U$ is defined locally in terms of Cantor sets of trajectories segments $C \times \mathbb{R}$. By definition, each trajectory segment defines a cosmological space-time $\mathcal M_U$ (`a world'). In this respect, a neighbouring trajectory on $I_U$ does not define `another world', but merely defines $U$ (`our world') at some earlier or later cosmic aeon. Using Bohmian language, one could say that such worlds are close relative to the implicate order defined by the p-adic metric on $I_U$ (see Section \ref{padic} below), but distant relative to the explicate order defined by the pseudo-Riemannian space-time metric. Importantly, individual trajectories of $U$ \emph{never} `branch' or `split' (as in the Everett Interpretation of quantum theory). 

A Cantor Set 
\be
\label{CN}
C=\bigcap_{k \in \mathbb{N}} C_k \nonumber
\ee
is defined as the intersection of its iterations $C_k$. The simplest Cantor set is the well-known ternary set $C(2)$ based on iterates of the unit interval on $\mathbb{R}$. For example, $C_k(2)$ is obtained by dividing each connected interval of $C_{k-1}(2)$ into 3 equal subintervals, and removing the middle open subinterval. More generally, let $C(p)$ denote a Cantor set where $C_k(p)$ is obtained by dividing each connected interval of $C_{k-1}(p)$ into $2p-1$ equal subintervals, and removing every second open subinterval and thus leaving $p$ pieces. The fractal similarity dimension of $C(p)$ is equal to $\log p/ \log (2p-1) \sim 1$ for large $p$. 

\begin{figure}
\centering
\includegraphics[scale=0.2]{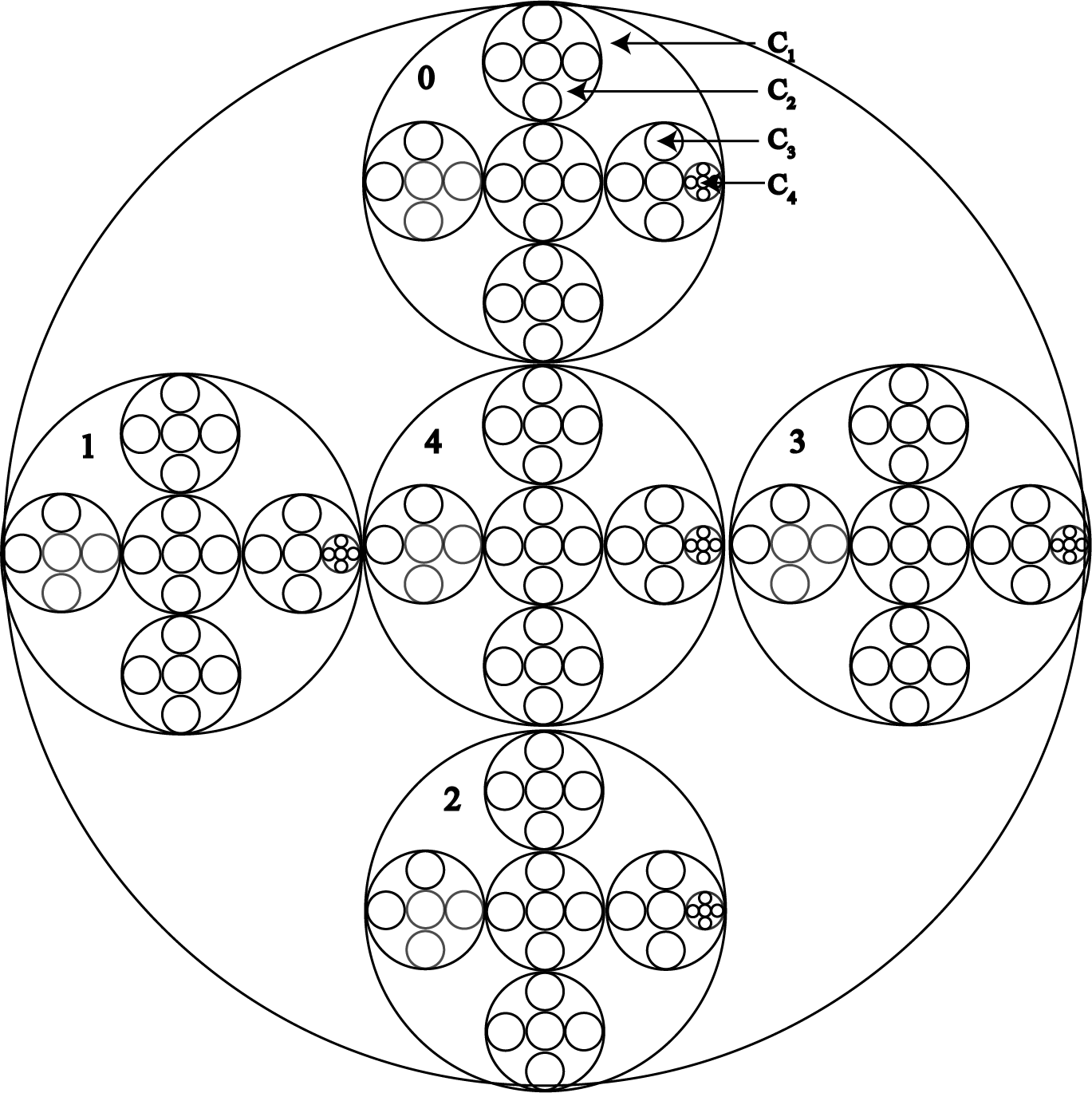}
\caption{A model for $C(5)$ - topologically equivalent to the 5-adic integers - based on disks at several levels of iteration.}
\label{padic}
\end{figure}

In Fig \ref{padic}, we illustrate an equivalent model for $C(5)$, whose elements are disks rather than intervals (2-balls rather than 1-balls). The zeroth iteration is the unit disk. For the first iteration, create 5 copies of the disk, each shrunk in area, and place inside the original disk as shown in the Figure. The second iteration is made by taking each disk of the first iteration, creating 5 shrunken copies of it and placing inside the disk as before. For the more general $p$, the $k$th iterate of $C(p)$ is created by taking $p$ shrunken copies of each disk of the $k-1$th iterate.  $C(p)$ is a topological model of the p-adic integers $\mathbb{Z}_p$ \cite{Robert}, discussed in Section \ref{padic}. 

\begin{figure}
\centering
\includegraphics[scale=0.5]{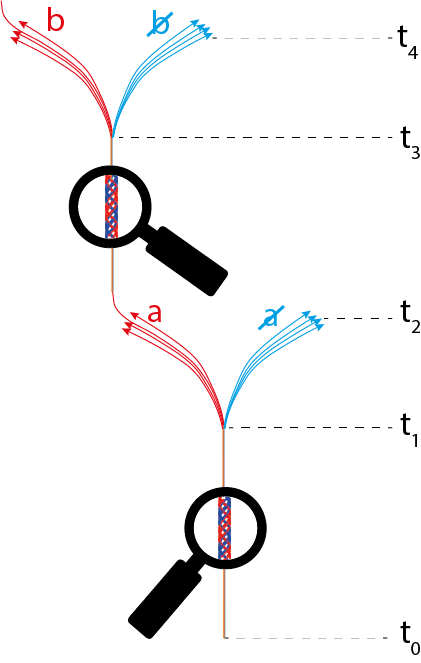}
\caption{A schematic illustration of state-space trajectories for the $k-1$th ($t_0<t<t_1$), $k$th ($t_1<t<t_3$) and $k+1$th ($t_3<t$) iterate trajectories of $\mathbb{R} \times C(p)$, with time varying instability, corresponding to a quantum system undergoing repeated measurement. The lower magnifying glass shows the $k$th iterate trajectories associated with a single $k-1$th iterate trajectory; the upper magnifying glass shows the $k+1$th iterate trajectories associated with a single $k$th iterate trajectory.}
\label{magglass}
\end{figure}

Fig \ref{magglass} shows schematically a fractal set of trajectory segments based on $\mathbb{R} \times C(p)$, at three levels of iteration ($k-1$, $k$ and $k+1$ for arbitrary $k$), and projected onto a two dimensional subset of state space. An evolving nonlinear dynamical systems will have temporally varying stability and predictability. This is manifest in Fig \ref{magglass}, showing regimes of metastability (with no divergence between neighbouring trajectories) punctured by intermittent periods of instability. Associated with each such instability, the trajectories are assumed attracted to two distinct quasi-stable regimes labelled `$a$' and `$\cancel{a}$', or `$b$' and `$\cancel{b}$' and trajectories bound for $a$ or $b$ are coloured red, whilst trajectories bound for $\cancel a$ or $\cancel b$ are labelled blue. The schematic is presumed to illustrate evolution on $I_U$ associated with repeated measurement of a quantum system. 

Between $t_0$ and $t_1$, Fig \ref{magglass} shows a single trajectory of the $k-1$ iterate of $\mathbb{R} \times C(p)$ where $p=2^N+1$. For simplicity, we refer to this as a `$k-1$th trajectory'. Under magnification (i.e. by zooming to the $k$th iterate), this trajectory is found to comprise a bundle of $2^N$ $k$th trajectories. These are coloured red or blue according to whether they are attracted (between $t_1$ and $t_2$) to the state-space regime $a$, or the state-space regime $\cancel a$. The fraction of trajectories attracted to $a$ can be written as  $n_k/2^N$ where $n_k \in \{0,1,....2^N\}$ takes one of $p$ values. (The fraction of trajectories attracted to $b$ is equal to $1-n_k/2^N$.) In this way, $n_k$ can be considered a label for one of the $p$ disks in the $k$th iterate of $C(p)$. 

Having evolved to the regime $a$, Fig \ref{magglass} shows a $k$th trajectory evolving through a second period of metastability between $t_2$ and $t_3$. Under magnification (i.e. by zooming to the $k+1$th iterate), this trajectory is found to comprise a bundle of $2^N$ $k+1$the trajectories. As before, a fraction $n_{k+1}/2^N$ of these are coloured red (and $1-n_{k+1}/2^N$ blue) according to whether they are attracted, between $t_3$ and $t_4$, to regime $b$, or regime $\cancel b$. 

In Invariant Set Theory, these quasi-stable regimes are presumed to correspond to measurement outcomes and the corresponding instability to the phenomenon of decoherence. Hence, as discussed further in Section \ref{hilbert}, the regions $a$ and $\cancel a$ are related to the eigenstates $|a\rangle$ and $|\cancel a\rangle$ (similarly for $b$) of quantum theory. The notion of eigenstates as symbolic representors of measurement outcomes is emphasised in Schwinger's approach to quantum theory \cite{Schwinger}. In Section \ref{gravity}, we consider the notion that the regimes $a$ and $\cancel a$ (and $b$ and $\cancel b$) are gravitationally distinct (or `clumped') regions of state space (c.f. \cite{Diosi:1989, Penrose:2004}). 

A key question is how to account for experimenter choice in making a measurement. Here one can note that the experimenter (and his or her neurons) are certainly part of $I_U$ and therefore both constrain and are constrained by dynamics on $I_U$. A crucial question is what  the structure of $I_U$ would have been had the experimenter chosen differently (in quantum mechanical language, what the outcome of measurements in a different basis to $|b\rangle$ and $|\cancel b\rangle$ would have been)? Resolution of the ontological status of these types of counterfactual experiments are absolutely central to Invariant Set Theory, and are discussed in detail in Section \ref{applications} in various settings.

A key characteristic of fractal invariant sets is their noncomputability \cite{Dube:1993, Blum}. One can interpret this as implying that there will be no computational rule for determining the outcome of a particular experiment, even though the underlying description of $I_U$ is completely deterministic. (In the case where $I_U$ is some quasi-fractal limit cycle, the corresponding result is that experimental outcomes cannot be estimated with a computational device smaller than $U$ itself). For this reason, the theory developed hereon is based on what in nonlinear dynamical systems theory is referred to as `symbolic dynamics' \cite{Williams}, implying that the quantitative aspect of the analysis below focusses on the topological characteristics of $I_U$.

\subsection{Complex Hilbert Vectors}
\label{hilbert}

Complex numbers play an essential role in quantum theory. In Invariant Set Theory, the structure of the sample space of $2^N$ trajectories (associated with a single $k-1$th trajectory) is guided by a permutation/negation representation of the complex roots of unity. Consider the bit string
\be
\label{bitstring}
S_a=\{a_1, a_2, a_3, a_4, \ldots a_{2^N}\}
\ee
where $a_i \in \{a, \cancel a\}$. That is to say, each of the $2^N$ $k$th iterate trajectory is labelled by the region $a$ or $\cancel a$ to which it is eventually attracted. We now define  two operators acting on $S_a$. The first is the cyclic operator 
\be
\label{zeta}
\zeta S_a= \{a_3, a_4, a_5, a_6, \ldots  a_{2^N}, a_1, a_2\}
\ee
which permutes the elements around the bit string in pairs. That is, $\zeta^{2^{N-1}} (S_a)=S_a$. The second is the permutation/negation operator  
\be 
\label{i}
i(S_a)=\{\cancel a_2, a_1, \; \cancel a_4, a_3 \; \ldots \;  \cancel a_{2^N}, a_{2^N-1}\}
\ee
which operates on pairs of elements and satisfies $i(i(S_a)\equiv i^2(S_a)=\{\cancel a_1, \cancel a_2, \cancel a_3, \cancel a_4  \ldots \cancel a_{2^N} \}\equiv -S_a$ and so $i^2(i^2(S_a)\equiv i^4(S_a)=S_a$. 

Now suppose $S_a$ has the form
\be
S_a=S_{a*}\|i(S_{a*})\|i^2(S_{a*})\|i^3(S_{a*})
\ee
where 
\be
S_{a*}=\{a, a,  a, \ldots, a\}
\ee
with $2^{N-2}$ elements and $\|$ is the concatenation operator. Then it is straightforwardly checked that 
\be
\label{zetai}
\zeta^{2^{N-3}}(S_a)=i(S_a) 
\ee
Taking roots of (\ref{zetai}) we have  
\be
\label{b}
\zeta (S_a)= i^{8/2^N}(S_a)
\ee
From the elementary theory of complex numbers
\be
\label{c}
e^{i \phi}=i^{2 \phi/\pi}
\ee
Hence, combining (\ref{b}) and (\ref{c}), we can write
\be
\label{zetan}
\zeta^n (S_a)= e^{i \phi} (S_a)\equiv S_a(\phi)
\ee
providing $2\phi/\pi=8n/2^{N}$, or
\be
\label{d}
\frac{\phi}{2\pi}=\frac{n}{2^{N-1}}
\ee
That is to say, it is necessary, in order to link these permutation/negation operators to complex numbers $e^{i\phi}$ and hence to angles $\phi$ on the circle, that $\phi$, as a fraction of $2 \pi$, can be described by $N-1$ bits (and hence can be described by $N$ bits). For other values of $\phi$, there is no correspondence between complex units and these permutation/negation operators. As discussed in Section (\ref{dirac}), the algebraic complex roots of unity arise from $\zeta$ only as a singular limit \cite{Berry} at $N=\infty$. 

For example, with $N=4$ 
\begin{align} 
S_a &= \{a,a, a, a,\ \ \cancel a, a, \cancel a,  a,\ \ \cancel a, \cancel a,  \cancel a,  \cancel a,\ \  a, \cancel a, a, \cancel a\} \nonumber \\
e^{i \pi/4} (S_a) \equiv  \zeta (S_a)&= \{a, a, \cancel a, a, \ \ \cancel a,  a, \cancel a, \cancel a, \ \ \cancel a,  \cancel a,  a, \cancel a, \ \ a, \cancel a,a,a\}=i^{1/2}(S_a)\nonumber \\
e^{i \pi/2} (S_a) \equiv  \zeta^2 (S_a)&= \{ \cancel a,  a,  \cancel a,  a, \ \ \cancel a, \cancel a, \cancel a, \cancel a, \ \ a, \cancel a,  a,  \cancel a, \ \ a, a, a, a \}=i(S_a) \nonumber \\
e^{i \pi} (S_a) \equiv  \zeta^4 (S_a)&= \{\cancel a,\cancel a, \cancel a, \cancel a,\ \ a, \cancel a, a, \cancel a,\ \ a, a, a, a,\ \  \cancel a,  a, \cancel a, a\}= i^2(S_a) \nonumber
\end{align}
Notice that $S_a$ is partially correlated with $e^{i \pi/4} (S_a)$, uncorrelated with $e^{i \pi/2} (S_a)$ and anti-correlated with $e^{i \pi} (S_a)$. 

The strings $S_a(\phi)$ will be used to define a sample space for complex Hilbert vectors. Recall that unit vectors in a real Hilbert space are a natural way to represent uncertainty about an object of study, even in classical physics. That is to say, if $\mathbf i$, $\mathbf j$ are orthogonal vectors, then, by Pythagoras's theorem, 
\be
\label{realhilbert}
\mathbf{v}(P_a)=\sqrt{P_a}\; \mathbf{i}+\sqrt{P_{\cancel{a}}} \;\mathbf{j} \nonumber
\ee
also has unit norm, for any probability assignments $P_a$, $P_{\cancel a}$ of events $a$ and $\cancel a$ respectively, where  $P_a +P_{\cancel a}=1$. On this basis, the Hilbert vector 
\be
\sqrt{\frac{n_k}{2^N}}\;|a\rangle + \sqrt{1-\frac{n_k}{2^N}}\; |\cancel a \rangle
\ee
will be used to represent a state of uncertainty as to whether a trajectory, randomly chosen from a set of $2^N$ trajectories belonging to some $k-1$th trajectory of $\mathbb{R} \times C(p)$, is attracted to the quasi-stable $a$ regime, or to the quasi-stable $\cancel a$ regime. Equivalently, if we write (\ref{realhilbert}) in the standard qubit form
\be
\label{realhilbert2}
\cos\frac{\theta}{2}\; |a\rangle + \sin\frac{\theta}{2}\; |\cancel a\rangle
\ee
then $\cos^2\theta/2$ must be expressible as a rational number of the form $n/2^N$, where $0\le n\le 2^N$, i.e. $\cos \theta$ must be expressible in the form $n/2^{N-1}-1$. 

Using the permutation/negation operator $\zeta^2$ and its relationship to complex roots of unity, we extend this aspect of $\mathbb{R} \times C(p)$ to complex Hilbert vectors. To do this let 
\be
S_a(\theta, \phi)=S_a(\phi)
\ee
but where the first $2^{N-1} \cos \theta$ occurrences of $\cancel a$ in $S_a(\phi)$ are set equal to $a$ if $0 \le \theta \le \pi/2$, or where the first $-2^{N-1} \cos \theta$ occurrences of $a$ in $S_a(\phi)$ are set equal to $\cancel a$ if $\pi/2 \le \theta \le \pi$. This implies for example
\begin{align}
S_a(0,\phi)&=\{a, a, a,  \ldots a\} \nonumber \\
S_a(\pi/2, \phi)&=S_a(\phi) \nonumber \\
S_a(\pi, \phi)&= \{\cancel a, \cancel a, \cancel a, \ldots \cancel a\}
\end{align}
and the probability that an element, randomly chosen from $S_a(\theta, \phi)$, is an $a$, is equal to $\cos^2 \theta/2$. In this way, when $\phi/2\pi$ and $\cos^2 \theta/2$ are describable by $N$ bits, we let
\be
\label{1qubitreal}
|\psi_a(\theta, \phi)\rangle \equiv
\cos\frac{\theta}{2}\;|a\rangle + e^{i \phi} \sin\frac{\theta}{2} |\cancel{a}\rangle \ee
represent a sample space $S_a(\theta, \phi)$ of trajectories corresponding to an iterate of $\mathbb{R} \times C(p)$. Now just as the quantum state is invariant under a global phase transformation, so $S_a(\theta, \phi)$ should be considered a sample space of bits, without order. That is to say, the set of global phase transformations on $|\psi_a\rangle$ should be considered equivalent to the set of permutation of elements of the bit string $S_a(\theta, \phi)$. 

As discussed in Section \ref{heisenberguncertainty} below, the conditions that $\phi/2\pi$ and $\cos^2 \phi/2$ are describable by $N$ bits, is in general, mutually incompatible. This non-commensurateness provides a number theoretic basis for a key property of quantum observables: operator non-commutativity. 

No matter how large is $N$, the mapping between bit strings and complex Hilbert vectors is an injection. Complex Hilbert vectors where $\phi/2\pi$ and $\cos^2 \theta/2$  are irrational never correspond to bit strings. This has important physical implications. As discussed below, certain counterfactual states allowed in quantum theory, are forbidden in Invariant Set Theory. As such, the algebraically closed complex Hilbert Space of quantum theory (closed under both multiplication \emph{and} addition) arises as the singular (and not the smooth) limit of Invariant Set Theory when $N$ (and hence $p$) is set equal to $\infty$. As discussed by Berry \cite{Berry}, old theories are typically the singular limit of theories which replace them. 

Because of the correspondence between complex Hilbert vectors and sample spaces of trajectories on $I_U$, it can be seen that Invariant Set Theory is fundamentally a $\psi$-epistemic theory of fundamental physics. 

\subsection{P-adic Integers and Cantor Sets}
\label{padic}

\begin{quote}

`We [number theorists] tend to work as much p-adically as with the reals and complexes nowadays, and in fact it it best to consider all at once.' (Andew Wiles, personal communication 2015.)

\end{quote}

The discussion above, where an essential difference was made between numbers describable or not describable by $N$ bits, may seem like `fine tuning' (especially when we are considering very large values of $N$). However, as will become clear, the distinction between an experiment where a crucial parameter can be described by $N$ bits, and one when it cannot, is a distinction between two experiments in state space, not in space time. 

To be more explicit, the natural metric of space-time is the pseudo-Riemannian metric and we expect our physical theories to be continuous in this metric. Hence if two experiments are performed in the real world, one with parameter settings $\phi_1$, the other with parameter settings $\phi_2$, then we expect these experiments to give, statistically at least, the same results as $\phi_2 \rightarrow \phi_1$. However, the nature of quantum paradoxes is that they involve counterfactual experiments: experiments that might have been but weren't. These are separated from actual experiments in state space, not physical space-time. As before, we should require that as the state-space distance between these experiments tends to zero, the experiments should be statistically indistinguishable. However, what is an appropriate metric on state space? Ostrowsky's theorem \cite{Katok} states that every nontrivial norm on $\mathbb{Q}$ is either equivalent to the Euclidean norm or the p-adic norm. Here we argue that the metric induced by the p-adic norm, rather than the Euclidean norm, is the appropriate measure of distance in state space, transverse to state-space trajectories. This leads to a radically different perspective on the fine-tuning issue above.  

By way of introduction to the p-adic numbers, consider the sequence
\be
\{1, 1.4, 1.41., 1.414, 1.4142, 1.41421 \ldots\} \nonumber
\ee
where each number is an increasingly accurate rational approximation to $\sqrt 2$. As is well known, this is a Cauchy sequence relative to the Euclidean metric $d(a,b)=|a-b|$, $a$, $b \in \mathbb{Q}$. 

Surprisingly perhaps, the sequence
\be
\{1, 1+2, 1+2+2^2, 1+2+2^2+2^3, 1+2+2^2+2^3+2^4, \ldots\}
\ee
is also a Cauchy sequence, but  with respect to the ($p=2$) p-adic metric $d_p(a,b)=|a-b|_p$ where
\be
|x|_p=\left \{%
\begin{array} {ll}
p^{-\textrm{ord}_p x} &\textrm{if } x \ne 0 \\
0 &\textrm{if } x=0
\end{array}%
\right.
\ee
and
\be
\textrm{ord}_p x= \left \{%
\begin{array}{ll}
\textrm{the highest power of \emph p which divides \emph x, if } x \in \mathbb Z \\
\textrm{ord}_p a - \textrm{ord}_p b, \textrm{ if } x=a/b, \ \ a,b \in \mathbb Z, \ b \ne 0
\end{array}%
\right.
\ee
Hence, for example
\be
d_2(1+2+2^2, 1+2)=1/4,\ \ d_2(1+2+2^2+2^3, 1+2+2^2)=1/8
\ee
Just at $\mathbb{R}$ represents the completion of $\mathbb{Q}$ with respect to the Euclidean metric, so the p-adic numbers $\mathbb{Q}_p$ represent the completion of $\mathbb{Q}$ with respect to the p-adic metric. A general p-adic number can be written in the form
\be
\sum_{k=-m}^{\infty} a_k p^k
\ee
where $a_{-m} \ne 0$ and $a_k \in \{0,1,2, \ldots, p-1\}$. The so-called p-adic integers $\mathbb Z_p$ are those p-adic numbers where $m=0$.  

It is hard to sense any physical significance to $\mathbb Z_p$ and the p-adic metric from the definition above. However, they acquire relevance in Invariant Set Theory by virtue of their association with fractal geometry. In particular, the map $F_2: \mathbb Z_2 \rightarrow C(2)$ 
\be
F_2: \sum_{k=0}^{\infty} a_k2^k \mapsto \sum_{k=0}^{\infty} \frac{2a_k}{3^{k+1}} \textrm{ where } a_k \in \{0,1\}
\ee  
is a homeomorphism \cite{Robert}, implying that every point of the Cantor ternary set can be represented by a 2-adic integer. More generally, 
\be
F_p: \sum_{k=0}^{\infty} a_kp^k \mapsto \sum_{k=0}^{\infty} \frac{2a_k}{(2p-1)^{k+1}} \textrm{ where } a_k \in \{0,1, \ldots p-1\}
\ee
is a homeomorphism between $\mathbb Z_p$ and $C(p)$ To understand the significance of the p-adic metric, consider two points $a, b \in C(p)$. Because $F_p$ is a homeomorphism, then as $d(a,b) \rightarrow 0$, so too does $d_p(\bar a, \bar b)$ where $F(\bar a)=a, \ F(\bar b)=b$. On the other hand, suppose $a \in C(p)$, $b \notin C(p)$. By definition, if $b \notin C(p)$, then $\bar b \notin \mathbb Z_p$. Let us assume that $b \in \mathbb Q$. Then $\bar b \in \mathbb Q_p$. This implies that $d_p(\bar a,\bar b) \ge p$. Hence,  $d(a,b) \ll 1 \centernot \implies d_p(\bar a, \bar b) \ll 1$. In particular, it is possible that $d_p(\bar a, \bar b) \gg 0$, even if $d(a,b) \ll 0$. From a physical point of view, a perturbation which seems insignificantly small with respect to the (intuitively appealing) Euclidean metric, may be unrealistically large with respect to the p-adic metric, if the perturbation takes a point on $C(p)$ and perturbs it off $C(p)$. The p-adic metric somehow encompasses the primal ontological property of lying on the invariant set. The Euclidean metric, by contrast, does not. 

Let $g(x, x')$ denote the pseudo-Riemannian metric on space-time, where $x, x' \in \mathcal M_U$. By contrast let $g_p(\mathcal M_U, \mathcal M_U')$ denote a corresponding metric in $U$'s state space, transverse to the state-space trajectories. As above, we suppose that if $\mathcal M_U \in I_U$, then $g_p(\mathcal M_U, \mathcal M_U') \rightarrow 0$ only in the p-adic sense, ie. only if $\mathcal M_U' \in I_U$. Here, a space-time $\mathcal M$ is presumed to belong to $I_U$ only when the relevant parameters $\cos^2 \theta/2$ and $\phi/2\pi$ are of the form $n/2^N$. When the parameters are not of this form then the corresponding space-time $\mathcal M'$ will not lie on $I_U$. Typically the latter occurs when considering the types of counterfactual experiment needed to prove quantum no-go theorems. If these $\mathcal M'$ do not lie on $I_U$ and are p-adically distant from the corresponding $\mathcal M$, then they are not admissible as feasible practically realisable experiments and the sorts of conspiracy arguments used to argue against so-called superdeterminist approaches to describe quantum physics fail. In this sense, use of the p-adic metric in state space essentially nullifies such no-go theorems. 

It can be noted in passing that many of the tools of analysis: algebra, calculus, Fourier transforms, and indeed Lie group theory, can be applied to the set of p-adic numbers \cite{Robert}. In addition, there is a natural measure on $C(p)$, the (self-similar) Haar measure. This provides a simple way to define the notion of probability and relate it to the concept of frequency of occurrence (an issue which is problematic in quantum theory - see \cite{Wallace}). 

\subsection{Multiple Bit Strings}
\label{entangle}
In Section \ref{hilbert}, a correspondence was developed between a bit string $\{a_1, a_2, \ldots a_{2^N}\}$ (modulo order) where $a_i \in \{a, \cancel a\}$, and the complex Hilbert vector $|\psi_a\rangle$. It is natural to consider an extension of this to a correspondence between a pair of bit strings $\{a_1, a_2, \ldots a_{2^N}\}$, $\{b_1, b_2, \ldots b_{2^N}\}$, where $b_i \in \{b, \cancel b\}$, and the tensor product $|\psi_a\rangle \otimes |\psi_b\rangle= |\psi_{ab}\rangle$. 

To start, consider the three bit strings
\begin{align}
\label{one}
S_a(\theta_1, \phi_1) &=\{a_1, a_2, a_3 \ldots a_{2^N}\} \mapsto |\psi_{a} (\theta_1, \phi_1) \nonumber \\
S_b(\theta_2, \phi_2) &= \{b'_1, b'_2, b'_3 \ldots b'_{2^N}\} \mapsto |\psi_{b}(\theta_2, \phi_2) \rangle  \nonumber \\
S_b(\theta_3, \phi_3) &=\{b''_1, b''_2, b''_3 \ldots b''_{2^N}\} \mapsto |\psi_{b}(\theta_3, \phi_3) \rangle  
\end{align}
where $b'_i, \; b''_i \in \{b, \cancel{b}\}$ and
\begin{align}
\label{g}
|\psi_a(\theta_1, \phi_1) & = \cos\frac{\theta_1}{2}|a\rangle + e^{i \phi_1} \sin\frac{\theta_1}{2}  |\cancel a\rangle \nonumber \\
|\psi_b(\theta_2, \phi_2) & = \cos\frac{\theta_2}{2}|b\rangle + e^{i \phi_2} \sin\frac{\theta_2}{2}  |\cancel b\rangle \nonumber \\
|\psi_a(\theta_3, \phi_3) & = \cos\frac{\theta_3}{2}|b\rangle + e^{i \phi_3}\sin\frac{\theta_3}{2}  |\cancel b\rangle.
\end{align}
Here we assume that the bit string $\{a_1 a_2 \ldots a_{2^N}\}$ is independent of $\{b'_1, b'_2, b'_3 \ldots b'_{2^N}\}$ and $\{b''_1, b''_2, b''_3 \ldots b''_{2^N}\}$. The three bits strings are reduced to two by setting
\begin{align}
\label{condition}
b_i&=b'_i \;\; \mathrm{if} \;\; a_i=a \nonumber \\
b_i&=b''_i \;\; \mathrm{if} \;\; a_i = \cancel{a}
\end{align}
whence
\be
\left.
\begin{array}{c}
\{a_1, a_2, a_3 \ldots a_{2^N}\}\nonumber\\
\{b_1, b_2, b_3 \ldots \;b_{2^N}\}
\end{array}
\large\right\} \mapsto |\psi_{ab}\rangle
\ee
where we write $|\psi_{ab}\rangle$ as
\begin{align}
\label{2qubit2}
|\psi_{ab}\rangle&=
\cos \frac{\theta_1}{2}|a\rangle
|\psi_b(\theta_2, \phi_2)\rangle
+e^{i\phi_1} \sin \frac{\theta_1}{2} |\cancel{a}\rangle |\psi_b(\theta_3,\phi_3)\rangle \\ \nonumber 
\end{align}
equivalent to the general form 
\be
\label{2qubit1}
|\psi_{ab}\rangle= \gamma^2_0 |a\rangle |b \rangle + \gamma^2_1 e^{i \chi_1} |a\rangle |\cancel{b} \rangle + \gamma^2_2 e^{i \chi_2}|\cancel{a}\rangle |b \rangle + \gamma^3_1 e^{i \chi_3}|\cancel{a}\rangle |\cancel{b} \rangle 
\ee
of a 2-qubit state, where
\begin{align}
\label{gamma}
\gamma_0&=\cos\frac{\theta_1}{2}\cos\frac{\theta_2}{2} \ \ \ \ 
\gamma_1=\cos\frac{\theta_1}{2} \sin\frac{\theta_2}{2} \ \ \ \ 
\gamma_2=\sin\frac{\theta_1}{2}\cos\frac{\theta_3}{2}  \ \ \ \ 
\gamma_3=\sin\frac{\theta_1}{2}\sin\frac{\theta_3}{2}  \nonumber \\
\chi_1&=\phi_2\ \ \ \ 
\chi_2=\phi_1 \ \ \ \ \
\chi_3=\phi_1+\phi_3 
\end{align}
It can be noted that there are six degrees of freedom associated with the two bit strings,  $\{a_1, a_2, b_3, \ldots a_{2^N}\}$,  $\{b_1, b_2, b_3 \ldots b_{2^N}\}$ consistent with quantum theory. Note also that these bit strings cannot in general both be written in the form $S_a(\theta, \phi)$,  $S_b(\theta, \phi)$, consistent with the fact that the state space $\mathbb S^6$ of the general 2-qubit cannot be expressed as the Cartesian product $\mathbb S^2 \times \mathbb S^2$ of Bloch Spheres. Based on (\ref{condition}), it is easy to show that statistical relations between the bit strings $\{a_i\}$ and $\{b_i\}$ are consistent with those from quantum theory. For example, the probability that  $a_i=a$ is equal to $\cos^2 \theta_1/2$ from (\ref{one}). Now if $a_i=a$ then by definition $b_i=b'_i$ from (\ref{condition}). The probability that $b'_i=b$ is equal to $\cos^2 \theta_2/2$ from (\ref{one}). Hence the probability that  $a_i=a$ and $b_i=b$ is equal to $\cos^2 \theta_1/2 \;\cos^2 \theta_2/2=\gamma^2_0$ from (\ref{gamma}).

Consider two special cases. Firstly, if $b'_i=b''_i$ for all $i$, then the probability that $a_i=a$ is independent of the probability that $b_i=b$. In other words, we can write
\be
\left.
\begin{array}{c}
S_a(\theta_1, \phi_1) \\
S_b(\theta_2, \phi_2)
\end{array}
\right\} \mapsto |\psi_{ab}\rangle
\ee
consistent with the factorisation $|\psi_{ab}\rangle=|\psi_a\rangle |\psi_b\rangle$ in quantum theory. In the second special case, let $b'_i=\cancel b''_i$ for all $i$ (so that $\cos^2\theta_3/2=\sin^2\theta_2/2$). Then the probability that $a_i=a$ and $b_i=b$ is equal to $\cos^2 \theta_1/2 \cos^2 \theta_2/2$, and the probability that $a_i= \cancel{a}$ and $b''_i=\cancel{b}$ is equal to $\sin^2\theta_1/2\sin^2\theta_3/2=\sin^2\theta_1/2\cos^2\theta_2/2$. Hence,  the probability that either $a_i=a$ and $b_i=b$, or $a_i= \cancel{a}$ and $b_i=\cancel{b}$ (i.e. the labels `agree') is equal to $\cos^2\theta_1/2 \cos^2 \theta_2/2+\sin^2\theta_1/2 \cos^2 \theta_2/2= \cos^2\theta_2/2$. With $\cos^2 \theta_1/2=1/2$, the correlation between $\{a_1, a_2, \ldots a_{2^N}\}$, $\{b_1, b_2, \ldots b_{2^N}\}$ are consistent with quantum theoretic correlations of measurement outcomes on the Bell state
\be
\label{bellstate}
|\psi_{ab} \rangle= \frac{|a\rangle|b\rangle+ | \cancel{a} \rangle | \cancel{b} \rangle}{\sqrt 2}
\ee
where $\theta_2$ denotes the relative orientation of EPR-Bell measurement apparatuses. The constraint that $\cos^2\theta/2$ is describable by $N$ bits takes centre stage in Invariant Set Theory's account of the Bell Theorem, as discussed below. 

This entanglement construction is generalised to $m$ qubits in Appendix \ref{MQ}. 

\subsection {Evolution on $I_U$ and the Dirac Equation}
\label{dirac}

In this Section we discuss how unitary dynamical evolution is described in Invariant Set Theory. We start by describing evolution of a uniformly moving particle of mass $m$ in a space-time $\mathcal M_U$. It will be convenient to discuss special relativistic theory from the outset. 

With suggestive notation, let $\psi(0)$ denote a set of four $2^N$ element bit strings witten in the form
\be
\psi(0)=
\bp S_1 \\ S_2 \\ S_3 \\S_4 \ep
\ee
To start, consider a frame where the particle is at rest. The time evolution of $\psi(0)$ is defined as
\be
\label{evolution1}
\psi(n\Delta t)=
\bp \zeta^n & \  & \  & \  \\\ & \zeta^n  & \  & \  \\ \  & \  & \zeta^{-n}  & \  \\ \  & \  & \  & \zeta^{-n}  \ep
\bp S_1 \\ S_2 \\ S_3 \\S_4 \ep
\ee
where $\Delta t = 2\pi /2^{N-1} \omega$, $\hbar \omega = mc^2$ and blank matrix elements denote zeros. From (\ref{zetan}) we can write
\be
\zeta^n \equiv e^{2 \pi i n/2^{N-1}}= e^{i \omega t}
\ee
where $t=n\Delta t$. Hence the set $\{e^{i \omega t}\}$ of time evolution operators is isomorphic to the multiplicative group of complex phases $\phi$ where $\phi / 2\pi$ is describable by $N-1$ bits. 
Writing the evolution equation (\ref{evolution1}) as
\be
\psi(t)=
\bp e^{i \omega t} & \  & \  & \  \\\ & e^{i \omega t}  & \  & \  \\ \  & \  & e^{-i \omega t}  & \  \\ \  & \  & \  & e^{-i \omega t}  \ep
\psi(0) \nonumber
\ee
the larger is $N$ the larger the number of phases $\phi$ where $e^{i \omega t}$ is defined. However, for any finite $N$, $\{e^{i \omega t}\}$ is not closed under addition. As discussed above, this is considered a desirable property of Invariant Set Theory, making it counterfactually incomplete. 

In the singular limit $N=\infty$ (or $1/N=0$), $e^{i \omega t}$ can be identified with the familiar complex exponential function, in which case, $e^{i \omega \Delta t} \approx 1+ i \omega \Delta t$ for small $\Delta t$ and
\be
i\partial_t \ e^{i \omega t}+ \omega e^{i \omega t}=0 \nonumber
\ee 
Because the limit is singular, the derivative is undefined for any finite $N$, no matter how big. In this sense, the Dirac equation for a particle at rest,
\be
\label{Dirac1}
i\hbar \gamma_0\; \partial_t \psi + mc \psi = 0
\ee
can be treated as the singular limit of (\ref{evolution1}) at $N=\infty$. Of course, in (\ref{Dirac1}), $\psi$ is to be considered some more abstract `wavefunction', lying in complex Hilbert Space. However, crucially, it should be noted that \ref{evolution1} is valid and well defined for finite $N$. That is to say, from the perspective of Invariant Set Theory, (\ref{evolution1}) should be considered more fundamental than (\ref{Dirac1}).

We now equate the energy $E$ of the particle in $\mathcal M_U$ with a key property of the geometry of the surrounding $k$th iterate trajectories on $I_U$: its periodicity. That is to say, writing $E=\hbar \omega$, then, because, $\hbar \omega = mc^2$, (\ref{evolution1}) is consistent with the special relativistic formula $E=mc^2$ for a particle at rest. What is the physics behind this construction? As discussed in the Introduction, the key idea in Invariant Set Theory is that the most primitive expression of the laws of physics is a description of the geometry of the invariant set $I_U$ in state space. That is to say, properties of our space-time $\mathcal{M}$ are determined by the geometry of $I_U$ in state space, in the neighbourhood of $\mathcal{M}$. Let us express the equation $E=\hbar \omega$ as an example of the following more general form
\be
\label{generalisedequation}
\parbox{2in}{Expression for a physical property in the locally Euclidean space $\mathcal M_U$}
\ \ 
=
\ \ \ 
\parbox{2in}{Expression based on the locally p-adic space $I_U$ }
\ee
This is analogous to the geodesic equation in general relativity written as
\be
\label{geodesic}
\frac {du^{\rho}}{d \lambda}= -\Gamma^{\rho}_{\mu\nu} u^{\mu}u^{\nu}
\ee
where the left hand side is a physical property of a test particle (its acceleration) on a geodesic and the right hand side involves derivatives transverse to the geodesic in space time. A key difference between (\ref{generalisedequation}) and (\ref{geodesic})  is that whilst calculus in $\mathcal M_U$ is based on the pseudo Riemannian metric, the calculus on $I_U$ must be based on the locally p-adic metric. Expression (\ref{generalisedequation}) is used to propose in Section \ref{gravity} a generalisation of general relativity, consistent with Invariant Set Theory. 

It can be noted that (\ref{evolution1}) describes two pairs of helical trajectories, rotating with opposite helicity. These can be associated with the two Weyl spinors of Dirac theory, and correspond to `particles' and `antiparticles' in $\mathcal M_U$. This linkage is again a manifestation of the basic notion expressed in (\ref{generalisedequation}).  

In a non-rest frame, we generalise (\ref{evolution1}) to 
\be
\label{evolution3}
\psi(t)= E_0E_1E_2E_3
\bp S_1 \\ S_2 \\ S_3 \\S_4 \ep
\ee
where
\begin{align}
&E_0=
\bp e^{i \omega t} & \  & \  & \  \\\ & e^{i \omega t}  & \  & \  \\ \  & \  &e^{-i \omega t}  & \  \\ \  & \  & \  & e^{-i \omega t}  \ep \ \ 
&E_1=
\bp \  & \  & \ & e^{i k_1 x} \\ \ & \   & e^{i k_1 x} & \  \\ \  & e^{-i k_1 x} &\  & \  \\ e^{-i k_1 x} & \  & \  & \  \ep \nonumber \\
&E_2=
\bp \  & \  & \ & ie^{-i k_2 y} \\ \ & \   & ie^{i k_2 y} & \  \\ \  & ie^{i k_2 y} &\  & \  \\ ie^{-i k_2 y} & \  & \  & \  \ep  \ \ 
&E_3=
\bp \  & \  & e^{i k_3 z} & \  \\ \ & \  & \  & e^{-i k_3 z} \\ e^{-i k_3 z} & \  & \ & \  \\ \  & e^{i k_3 z} & \  &\  \ep \nonumber
\end{align}
where $x=n \Delta x$, $\Delta x = 2 \pi/2^{N-1} k_1$ (etc for $y$ and $z$) and $\hbar^2 \omega^2= \hbar^2 |\mathbf k |^2+m^2$ (leaving three degrees of freedom to describe the particle's momentum in $\mathcal M_U$). Note that the Dirac matrices
\begin{align}
\ \nonumber \\
&\gamma_0=
 \bp \; 1&\ &\ &\  \\  \ &1 &\ &\   \\   \ & \ &-1 &\  \\  \ &\ &\ &-1  \ep 
\ \ 
&\gamma_1=
 \bp \; \ &\ &\ &1 \\  \ &\ &1&\   \\   \ &-1 &\ &\  \\  -1&\ &\ &\  \ep \nonumber \\ 
 \ \nonumber \\
 &\gamma_2=
 \bp \; \ &\ &\ &-i \\  \ &\ &i&\   \\   \ &i &\ &\  \\  -i&\ &\ &\  \ep 
\ \ 
&\gamma_3=
 \bp \; \ &\ &1&\  \\  \ &\ &\ &-1   \\   -1&\  &\ &\  \\  \ &1 &\ &\  \ep  \nonumber
\end{align}
are clearly evident in the form of the evolution operators $E_0$, $E_1$, $E_2$ and $E_3$. With particle momentum $\mathbf p$ in $\mathcal M_U$ inherited from the periodicity of the surrounding geometry of $I_U$, so that $\mathbf p = \hbar \mathbf k$ c.f. (\ref{generalisedequation}), then (\ref{evolution3}) implies that
\be
E^2=p^2+m^2 \nonumber
\ee
as required by special relativity theory. 

As before, in the singular limit $N=\infty$ (and only in this limit) we can write $i\partial_x e^{ik_1x} + k_1e^{i k_1x}=0$ and so on. Hence, (\ref{evolution3}) can be seen to be equivalent to evolution as described by the full Dirac equation
\be
\label{Dirac2}
i\hbar\gamma_i\; \partial_i \psi + mc \psi = 0
\ee
in the (singular) limit $N=\infty$. As an evolution equation for finite $N$, (\ref{evolution3}) is to be considered (in Invariant Set Theory) as more fundamental than (\ref{Dirac2}). Like the Dirac equation, (\ref{evolution3}) is relativistically invariant. Interestingly, the evolution (\ref{evolution3}) in space and time is fundamentally granular. The implications of this will be explored elsewhere. 

Equation (\ref{evolution3}) describes the periods of metastable evolution on $I_U$ as shown schematically in Fig \ref{magglass} (between $t_0<t<t_1$ and between $t_2<t<t_3$), i.e. corresponding to unitary evolution between preparation and measurement in quantum theory. In Section \ref{gravity}, the periods of instability and attraction to regions $a$ and $\cancel a$ (between $t_1<t<t_2$) and $b$ and $\cancel b$ (between $t_3<t<t_4$) are linked to explicitly gravitational processes. 

\section{Applications of Invariant Set Theory}
\label{applications}

\subsection{The Bell Theorem}
\label{belltheorem}

Invariant Set Theory is both realistic, and, as discussed below, locally causal. Despite this, it robustly violates the Bell inequalities without fine-tuned conspiracy, retrocausality or denial of experimenter free will. The CHSH \cite{CHSH} version of the Bell Inequality can be written in the form
\begin{equation}
\label{chsh}
|\mathrm{Corr}(A_1, B_1) - \mathrm{Corr}(A_1, B_2)|+|\mathrm{Corr}(A_2, B_1)+\mathrm{Corr}(A_2, B_2)| \le 2
\end{equation}
Here Alice and Bob each choose one of two buttons labelled `1' and `2' to perform spin measurements on entangled particle pairs prepared in the Bell state (\ref{bellstate}). Here `Corr' denotes the correlation between spin measurements over some ensemble of entangled particle pairs. According to both Quantum Theory and Invariant Set Theory, $\mathrm{Corr}(A_1, B_1)=\cos \theta_{A_1B_1}$, where $\theta_{A_1B_1}$ denotes the relative orientation of the measuring apparatuses when Alice presses button 1, Bob presses button 2 etc. With $A_1$, $A_2$ and $B_1$ collinear (not a vital assumption; see \cite{Palmer:2015, Palmer:2015b}) and $\theta_{A_1B_1}=\theta_{A_1A_2}+\theta_{A_2B_1}$, then we can write
\be
\label{cosab}
\cos \theta_{A_1B_1}=\cos \theta_{A_1A_2} \cos \theta_{A_2B_1} - \sin\theta_{A_1A_2} \sin\theta_{A_2B_1}.
\ee
Now let us assume that $\cos \theta_{A_2B_1}$ can be written in the form $n_1/2^N$. This will be the case in Invariant Set Theory over the subsample of measurements where Alice choses button 2 and  Bob button 1. We can also assume $\cos \theta_{A_1A_2}$ is in the form $n_2/2^N$ since Alice can, having measured her particle relative to $A_2$, remeasure the same particle relative to $A_1$ - such a re-measurement experiment must also be described by a state on the Invariant Set. However, then $\sin \theta_{A_2B_1}$ and $\sin \theta_{A_1A_2}$ cannot be written in the form $n_3/2^N$ and $n_4/2^N$ - if they could then there would be integer solutions to $n_1^2+n_3^2=n_2^2+n_4^2=2^N$ (we assume none of $n_1$, $n_2$, $n_3$ or $n_4$ is identically zero). As first shown by Euclid, there are no non-zero integer solutions to the Pythagorean equations $a^2+b^2=c^2$ with $c$ a power of 2. With $\theta_{A_1A_2}$ and $\theta_{A_2B_1}$ being independent angles, i.e. with no functional relationship, (\ref{cosab}) implies that $\cos\theta_{A_1B_1}$ is not describable by $N$ bits (no matter how large is $N$). 

However, if $\cos\theta_{A_1B_1}$ is not describable by $N$ bits, then the hypothetical experiment where Alice presses $A_1$ and Bob $B_1$ on the particle pair where Alice actually presses $A_2$ and Bob $B_1$, does not lie on $I_U$. That is to say, if Alice chooses button 2 and Bob button 1, then the counterfactual world where, for the same entangled state, Alice chooses button 1 and Bob button 1, does not correspond to a state of the universe on $I_U$. Similarly, the counterfactual world where Alice chooses button 2 and Bob button 2 does not lie on $I_U$. Conversely, if Alice chooses button 1 and Bob button 1 (in $\mathcal M_U \in I_U$), then the counterfactual worlds where, for the same entangled state, Alice chooses button 1 and Bob button 2, or Alice button 2 and Bob button 1, do not correspond to states of the universe on $I_U$. Hence it is impossible to derive the CHSH inequality on any single sample of particle pairs in Invariant Set Theory. 

Does the Invariant Set constraint - that the cosine of relative orientation between Alice and Bob's apparatuses are describable by $N$ bits - violate local causality? No. Let $E_A$ and $E_B$ denote the setting of Alice and Bob's measuring apparatus, assumed space-like separated. The necessarily finite precision of these apparatuses mean that if $N$ is large enough, then whilst Alice and Bob together have complete control over the leading bits of the cosine of their relative orientation, they have no control over the trailing bits, and in particular have no control over whether or not the cosine of relative orientation is describable by $N$ bits. Since the choice of leading bits is irrelevant to the issue of whether or not the relative orientation is describable by $N$ bits, Alice's choice of setting for the orientation of her measuring apparatus does not in any way constrain Bob's choice in setting his apparatus: $E_A$ does not affect $E_B$ or \emph{vice versa}. Rather, the uncontrollable constraint that the cosine of the relative orientation must be describable by $N$ bits is a holistic property of the Invariant Set. Just as the causal ordering of events in Special Relativity is not destroyed by the presence of space-time curvature (and hence distant masses) in General Relativity, so neither does the geometry of $I_U$ in state space destroy the causal structure of General Relativistic space-time: the holistic nature of the Invariant Set postulate is not non-causal. Although, the Invariant Set postulate technically implies a violation of the Measurement Independence condition \cite{Hall:2010, Hall:2011}, at a deeper level it indicates the limitation of conventional reductionist approaches to the formulation of physical theory: in Invariant Set Theory the physics of the small is determined by the large-scale state-space structure of the universe. 

So what is actually measured when the CHSH inequality is shown to be violated experimentally? It is important to note that whilst the derivation of the CHSH inequality assumes a single sample of particles over which different spin measurements are performed, in a real-world experiment each of the four correlations in the CHSH inequality are estimated from four sub-experiments using separate sub-ensembles of entangled particle pairs. In Invariant Set Theory, each sub-experiment takes places in $\mathcal M$ on $I_U$. This means that experimentally the four relative angles for each of the four sub-experiments cannot be \emph{precisely} the four angles $\theta_{A_1B_1}$, $\theta_{A_1B_2}$, $\theta_{A_2B_1}$ and $\theta_{A_2B_2}$ above, but must rather be, say, $\theta'_{A_1B_1}$, $\theta_{A_1B_2}$, $\theta_{A_2B_1}$ and $\theta'_{A_2B_2}$ where the cosine of each angle is describable by $N$ bits. Here we recognise the finite precision of real-world experiments. In particular, both $|\theta'_{A_1B_1}-\theta_{A_1B_1}|$ and $|\theta'_{A_2B_2}-\theta_{A_2B_2}|$ can be assumed smaller than the finite precision within which these experiments can be performed.

Even though $|\theta'_{A_1B_1}-\theta_{A_1B_1}|\ll1$, the argument above does not require the experimenters to be in any way precise in setting up quantum no-go experiments, i.e. in ensuring that the relative orientation is $\theta'_{A_1B_1}$ rather than $\theta_{A_1B_1}$. It is worth discussing this in the context of Bell's remark \cite{Bell}:
\begin{quote}
Remember, however, that the disagreement between locality and quantum mechanics is large \ldots. So some hand trembling [by the experimenters] can be tolerated without much change in the conclusion. Quantification of this would require careful epsilonics.
\end{quote}
Since by construction $\theta_{A_1B_1}$ defines a state which does not lie on $I_U$, and $\theta'_{A_1B_1}$ defines a state that does lie on $I_U$, the p-adic distance between these states is large, even though the Euclidean distance between the angles is small. In particular, with respect to the p-adic metric, which we claim to be the physically relevant metric transverse to trajectories in state space, there is no small perturbation (e.g. associated with experimenter hand trembling) which can take a state of the universe from one where the relative orientation of the measuring apparatuses is $\theta'_{A_1B_1}$ to one where the relative orientation is $\theta_{A_1B_1}$. Hence we conclude that application of `careful epsilonics' in fact allows us to conclude that CHSH violation is robustly consistent with local realism, even though the argument does appear fine-tuned and hence conspiratorial with respect to the physically less relevant Euclidean metric. This discussion is central to all so-called conspiracy arguments \cite{Bell} concerning the Bell inequalities: essentially there are no conspiracies if one measures distances in state space p-adically. 

One can phrase the issue in terms of robustness to noise. One can perturb a state on $I_U$ with small amplitude noise. However, for the noise to be physical and hence respect the Invariant Set postulate, then the noise must be added p-adically to an unperturbed state (consistent with the property that the set of p-adic integers is closed under addition). That is to say, one can p-adically perturb a state where $\theta'_{A_1B_1}$ is describable by $N$ bits with small amplitude noise and the relative orientation $\theta''_{A_1B_1}$ of the perturbed state will still be describable by $N$ bits. Conversely, no amount of such noise will take the state where $\theta'_{A_1B_1}$ can be described by $N$ bits, to one where $\theta_{A_1B_1}$ is not described by $N$ bits - the states are too far apart for this to be possible. 

To conclude this Section, it is worth noting that the argument given here is not incompatible with a recent discussion by 't Hooft \cite{tHooft:2015} (based on a model \cite{tHooft:2015b} which, like Invariant Set Theory, is deterministic, nonlinear, and eschews superpositions at a fundamental level): 

\begin{quote}

\ldots even in a superdeterministic world, contradictions with Bell's theorem would ensue if it would be legal to consider a change of one or a few bits in the beables describing Alice's world, without making any modifications in Bob's world. \ldots [However,] it is easy to observe that, certainly in the distant past, the effects of such a modification would be enormous and it may never be compatible with a simple low-entropy Big Bang \ldots Thus, we can demand in our theory that a modification of just a few beables in Alice's world without any changes in Bob's world is fundamentally illegal. This is how an ontological deterministic model can `conspire' to violate Bell's theorem. 

\end{quote}
Somewhat similar to Invariant Set Theory, this quote proposes a constraint on the physics of the small from the physics of the large. On the other hand, the supposed conflict with the low entropy state of the initial universe is speculative, and is not (as far as the author can see) a direct consequence of the cellular automaton dynamics \cite{tHooft:2015b} proposed by 't Hooft. By contrast, in Invariant Set Theory, the illegality of modifying Alice's world is a direct and immediate consequence of the assumed Invariant Set postulate. There are other differences with 't Hooft's theory. For example, although 't Hooft's model is not founded on differential evolution equations, it is nevertheless based on a finite-difference cellular automaton rule for dynamical evolution. That is to say, dynamical evolution falls into the generic class of `hidden-variable' theories. In particular, its underpinning metaphysics differs conceptually from the non-reductionist Invariant Set Theory philosophy.

\subsection{Quantum Interferometry}
\label{heisenberguncertainty}

\begin{figure}
\centering
\includegraphics[scale=0.4]{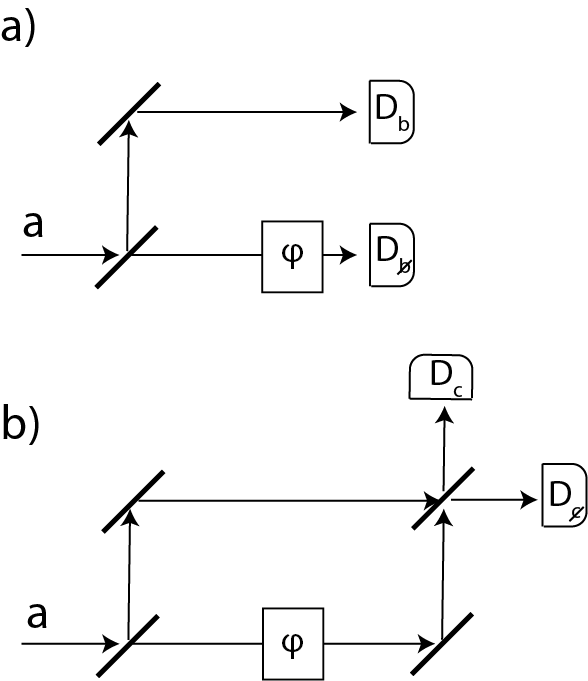}
\caption{a) A which-way experiment. b) A interferometric experiment}
\label{MachZehnder}
\end{figure}

It is now possible to show how the non-commutativity of position and momentum observables in quantum theory is explainable in Invariant Set Theory through the number theoretic incommensurateness between $\phi$ and $\cos \phi$. Consider the experimental set up as in Fig \ref{MachZehnder}a. According to quantum theory, the input state vector $|a\rangle$ is transformed by the unitary operators
\be
 U=\bp \; 1 &1 \\ 1& -1 \; \ep; \ \ \  V=\bp \; 1 &0 \\ & e^{i\phi} \; \ep \nonumber
 \ee
according to
\be
|a\rangle \stackrel{VU}{\mapsto} \frac{1}{\sqrt 2}(|b\rangle + e^{i \phi} |\cancel b\rangle)
\ee
According to (\ref{1qubitreal}), the corresponding transformation between preparation and measurement state on $I_U$ is
\be
\label{whichway}
S_a(0,0) \mapsto S_b(\pi/2,\phi)
\ee
Consider, instead, the experimental set up as in Fig \ref{MachZehnder}b (that is to say, a Mach-Zehnder interferometer). Now, according to quantum theory, the input state vector $|a\rangle$ is transformed according to
\be
|a\rangle \stackrel{UVU}{\mapsto} \cos\frac{\phi}{2}|c\rangle + \sin\frac{\phi}{2} |\cancel c\rangle
\ee
According to (\ref{1qubitreal}), the corresponding transformation on $I_U$ is
\be
\label{interfere}
S_a(0,0) \mapsto S_c(\phi,0)
\ee 
We now claim that transformations (\ref{whichway}) and (\ref{interfere}) are incompatible, in the sense that  if transformation (\ref{whichway}) corresponds to a trajectory on the invariant set $I_U$, then transformation (\ref{interfere}) does not, and \emph{vice versa}. To see this we need a result from elementary number theory:
\bigskip
$\mathbf{Theorem}$\cite{Jahnel:2005}.  Let $\phi/\pi \in \mathbb{Q}$. Then $\cos \phi \notin \mathbb{Q}$ except when $\cos \phi =0, \pm 1/2, \pm 1$. 
\\
$\mathbf{Proof}$. Assume that $2\cos \phi = a/b$ where $a, b \in \mathbb{Z}, b \ne 0$ have no common factors.  Since $2 \cos 2\phi = (2 \cos \phi)^2-2$ then
\be
2\cos 2\phi = \frac{a^2-2b^2}{b^2}
\ee
Now $a^2-2b^2$ and $b^2$ have no common factors, since if $p$ were a prime number dividing both, then $p|b^2 \implies p|b$ and $p|(a^2-2b^2) \implies p|a$, a contradiction. Hence if $b \ne \pm1$, then the denominators in $2 \cos \phi, 2 \cos 2\phi, 2 \cos 4\phi, 2 \cos 8\phi \dots$ get bigger without limit. On the other hand, if $\phi/\pi=m/n$ where $m, n \in \mathbb{Z}$ have no common factors, then the sequence $(2\cos 2^k \phi)_{k \in \mathbb{N}}$ admits at most $n$ values. Hence we have a contradiction. Hence $b=\pm 1$ and $\cos \phi =0, \pm1/2, \pm1$ QED. 
\bigskip

Suppose an experimenter performs a `which way' experiment as in Fig \ref{MachZehnder}a.  As before, the experimenter is assumed to have control only on the leading bits of $\phi$ and has no control on the trailing bits, or on the Invariant Set condition that $\phi/2\pi$ is describable by $N$ bits. It is exponentially unlikely (with $N$) that $\cos \phi$ would correspond precisely to one of the exceptional values in the theorem. Hence, in general, one can state that in Invariant Set Theory the number-theoretic incommensurateness between $\phi$ and $\cos \phi$ implies that if a position measurement is made, then a momentum measurement cannot be made, and \emph{vice versa}. 

Feynman famously claimed \cite{FeynmanHibbs} that quantum interference demonstrates that the Laplacian laws for combining probabilities necessarily fail in quantum physics. We can use the discussion above to argue otherwise. Let us assume $\phi \approx 0$, and let $P_c \approx 0$ denote the probability of detection by $D_c$ in Fig \ref{MachZehnder}b, $P_b$ the probability of detection by $D_b$ and $P_{\cancel b}$ the probability of detection by $D_{\cancel b}$ in Fig \ref{MachZehnder}a. In general, $P_c \ne P_b+P_{\cancel b}=1$. However, as discussed above, the notion of probability is defined from the natural Haar measure on $I_U$. As discussed above, in situations where momentum measurements are made ($\cos \phi$ describable by $N$ bits), counterfactual position measurements ($\phi/\pi$ describable by $N$ bits) do not lie on $I_U$. That is to say, the sample space from which the probability $P_c$ is estimated is completely distinct from the sample space where position measurements are made. As such, one can assert in Invariant Set Theory that particles do travel either through the upper of lower branch. However, if a momentum measurement is being made, then the positions of such particles are fundamentally unobservable - in Bell's language, their positions are merely `beables'. 

If particles do travel through either the upper or lower arm in situations were an interference experiment is being conducted, then a key `conceptual' question is: How does a particle `know' whether to behave like a wave rather than a classical particle? This question becomes potentially problematic when one considers a delayed-choice experiment, where the experimenter only decides whether to perform an interferometric or which-way experiment after the particle has entered the apparatus - and perhaps passed through the phase shifter. However, in Invariant Set Theory, this isn't problematic at all: the structure of $I_U$ at some time $t_0$ is determined by events to the future of $t_0$. This isn't to be confused by the notion of retrocausality, but is again indicative of the holistic causal nature of $I_U$. The situation is fundamentally no different to asking whether a point $p \in \mathcal M_U$ lies on a black hole event horizon \cite{Palmer:2015}, defined as the boundary of null rays which escape to (future null) infinity. Like $I_U$, the event horizon is an entirely causal and realistic entity. However, because of the global definition of the event horizon, it is impossible to determine its position from a knowledge of the Riemann tensor in the neighbourhood of the event horizon; its position at some time $t_0$ in space-time can be influenced by an event potentially in the future of $t_0$, e.g. whether or not a massive object falls into the black hole at $t_1>t_0$. As such, the position of the event horizon at $t_0$ can also depend on the state of some experimenter's neurons at some future time, if these neurons are responsible for determining whether or not the massive object falls into the black hole.   

At the practical level, the finite-precision with which an experimenter can set $\phi$ means that it is impossible to ascertain by experiment whether $\cos \phi$ is describable by $N$ bits, or $\phi/\pi$ is describable by $N$ bits. Moreover, at the computational level there is no algorithm for determining the precise value of $\phi$. More specifically, since the geometric properties $I_U$ are non-computable \cite{Blum, Dube:1993}, there is no algorithm for determining what type of experiment (which way or interferometric) will be performed ahead of time. Because of this, Invariant Set Theory explains naturally why, in quantum physics, statements about the future have to be cast in overtly probabilistic language, whilst statements about the past can be cast in quite definite non-probabilistic terms \cite{Dyson:2004}. 

\subsection{PBR}
\label{puseytheorem}

The recent PBR theorem \cite{Pusey} is a no-go theorem casting doubt on $\psi$-epistemic theories (where the quantum state is presumed to represent information about some underlying physical state of the system). Unlike CHSH where Alice and Bob each choose measurement orientations A or B, here Alice and Bob, by each choosing $0$ or $1$, prepare a quantum system in one of four input states to some quantum circuit: $|\psi_0\rangle |\psi_0\rangle$, $|\psi_0\rangle |\psi_1\rangle$, $|\psi_1\rangle |\psi_0\rangle$ or $|\psi_1\rangle |\psi_1\rangle$, where
\begin{align}
|\psi_0\rangle&=\cos \frac{\theta}{2} |0\rangle + \sin \frac{\theta}{2} |1\rangle \nonumber \\
|\psi_1\rangle&=\cos \frac{\theta}{2} |0\rangle - \sin \frac{\theta}{2} |1\rangle \nonumber
\end{align}
In addition to the parameter $\theta$, the circuit contains two phase angles $\alpha$ and $\beta$; as discussed below, the phase angle $\alpha$ most closely plays the role of the phase angle $\phi$ in the Mach-Zehnder interferometer in Section \ref{heisenberguncertainty}. The output states of the quantum circuit are characterised as `$\mathrm{Not}\; 00$', `$\mathrm{Not}\; 01$', `$\mathrm{Not}\; 10$' and `$\mathrm{Not}\; 00$'.  The $\alpha$ and $\beta$ are chosen to ensure that (according to quantum theory), if Alice and Bob's input choices are $\{IJ\}$ where $I,J\in\{0,1\}$, then the probability of `$\mathrm{Not}\; IJ$' is equal to zero. However, if physics is governed by some underpinning $\psi$-epistemic theory, then, so the argument goes, at least occasionally the measuring device will be uncertain as to whether, for example, the input state was prepared using $00$ and $01$ and on these occasions it is possible that an outcome `$ \mathrm{Not} \; 01$' is observed when the state was prepared as $01$, contrary to quantum theory (and experiment). How does Invariant Set Theory, which is indeed a $\psi$-epistemic theory, avoid this problem?

Working through the algebra, it is found that the probabilities of various outcomes are trigonometric functions of $\alpha-\beta$, $\alpha-2\beta$, $\beta$ and $\theta$. For example, if Alice and Bob chose $00$, then, according to quantum theory (and therefore experiment), the probability of obtaining the outcome `$\mathrm{Not}\; 01$' is equal to 
\be
\label{XXX}
X= \cos^4 \frac{\theta}{2} + \sin^4 \frac{\theta}{2} +2 \cos^2 \frac{\theta}{2} \sin^2 \frac{\theta}{2}\cos (\alpha-2\beta) \ne 0
\ee
On the other hand, if Alice and Bob chose $01$, then the probability of obtaining the outcome `$\mathrm{Not}\; 01$' would be equal to 
\be
\label{ZZZ}
Z=X-4 \cos^2 \frac{\theta}{2} \sin^2 \frac{\theta}{2}-4 \cos^3 \frac{\theta}{2} \sin \frac{\theta}{2}\cos(\alpha-\beta)-4 \cos \frac{\theta}{2} \sin^3 \frac{\theta}{2}\cos\beta=0
\ee
The key point is that $X$ contains the trigonometric term $\cos(\alpha-2\beta)$, whilst $Z$ contains the terms $\cos(\alpha-\beta)$ and $\cos \beta$. Now one can clearly find values for $\alpha$, $\beta$ and $\theta$ such that $X$ is described by $N$ bits. That is to say, for large enough $N$, Invariant Set Theory can predict the quantum theoretic probability of outcome `$\mathrm{Not}\;01$' when Alice and Bob chose 00. However, in general it is impossible to find values for these angles such that $X$ and $Z$ are \emph{simultaneously} describable by $N$ bits. The number-theoretic argument is exactly that used to negate the Bell Theorem. For example, if $\cos (\alpha-2\beta)$ and $\cos \beta$ are describable by a finite number of bits, then $\cos(\alpha-\beta)=\cos(\alpha-2\beta) \cos \beta+ \sin(\alpha-2\beta) \sin \beta$ is not. This means that if in reality Alice and Bob chose $00$ in preparing a particular quantum system and the outcome was `$\mathrm{Not}\; 01$' , then there is no counterfactual world on $I_U$ where Alice and Bob chose $01$ in preparing the same quantum system, and the outcome was again `$\mathrm{Not}\; 01$'. That is to say, it is not the case that $Z=0$ for this counterfactual experiment - rather, $Z$ is undefined. Conversely, if in reality Alice and Bob chose 01, then there exist values for $\alpha$, $\beta$ and $\theta$ such that $Z$ is described by $N$ bits and equal to zero (to within experimental accuracy) and `$\mathrm{Not}\; 01$' is not observed. 

Let $\{\alpha_X, \beta_X, \theta_X\}$ denote a set of angles such that $X$ is describable by $N$ bits, and $\{\alpha_Z, \beta_Z, \theta_Z \}$ a set of angles such that $Z$ is describable by $N$ bits, i.e. these correspond to experiments on $I_U$. Now, as before, we can find values such that the differences $\alpha_Z-\alpha_X$, $\beta_Z-\beta_X$ and $\theta_Z-\theta_X$ are each smaller than the precision by which these angles can be set experimentally. Hence, Invariant Set Theory can readily account for pairs of experiments performed sequentially with seemingly identical parameters, the first where Alice and Bob choose $00$ and the outcome is sometimes `$\mathrm{Not}\; 01$', and the second where Alice and Bob choose $01$ and the outcome is never `$\mathrm{Not}\; 01$'. That is to say, the Invariant Set Theoretic interpretation of the PBR quantum circuit reveals no inconsistency with experiment. Even though Invariant Set Theory is $\psi$-epistemic, the holistic structure of the invariant set $I_U$ ensures that the measuring device will never be uncertain as to whether, for example, the input state was prepared using $00$ and $01$.

It was shown above that Invariant Set Theory evades the Bell theorem by violating the Measurement Independence assumption. Here it has been shown that Invariant Set Theory evades the PBR theorem by violating an equivalent Preparation Independence assumption. As before a more revealing reason for the failure of these theorems is that Invariant Set Theory does not fit into the traditional reductionist approach to hidden-variable theory. As before, this does not conflict at all with the experimenter's sense of free will. Neither does it imply fine-tuning with respect to the physically relevant p-adic metric on $I_U$. 

\section{Quantum Gravity}
\label{gravity}

\begin{quote}
`Despite impressive progress \ldots towards the intended goal of a satisfactory quantum theory of gravity, there remain fundamental problems whose solutions do not appear to be yet in sight. \ldots [I]t has been argued that EinsteinÕs equations should perhaps be replaced by something more compatible with conventional quantum theory. There is also the alternative possibility, which has occasionally been aired, that some of the basic principles of quantum mechanics may need to be called into question.' (Roger Penrose, 1976 \cite{Penrose:1976}
\end{quote}
Using Invariant Set Theory as a guide, we speculate here on one such `alternative possibility', and in so doing propose a generalisation of Einstein's equations. This in turn suggests some potentially observable consequences of Invariant Set Theory. 

As discussed, quantum theory emerges from Invariant Set Theory as the singular limit where $1/N=0$. As Michael Berry has noted, old theories are typically the singular limits of new theories: classical physics and quantum theory, Newtonian gravity and general relativity being two pairs of relevant examples. As in both of these pairs of examples, the old theories often work well under limited parameter regimes, but can fail catastrophically. 

We speculate here that quantum theory may itself fail catastrophically under situations where gravity is not negligible. For example, it seems plausible to speculate that the relative stability of the two identifiable regions $a$ and $\cancel a$, and the corresponding instability of the basin boundary - both geometric properties of $I_U$ -  is a manifestation of the phenomenon we call `gravity'. This is consistent with the notion \cite{Diosi:1989} \cite{Penrose:2004} that measurement eigenstates in quantum theory are gravitationally distinct. If this is correct it will mean that gravity will itself be inherently `decoherent' - quite different from the effects of the other gauge fields. This idea should be testable experimentally in the coming years. 

How are the gauge fields representable in Invariant Set Theory? Instead of thinking of these as fields on a fixed background space-time, one should instead think them in terms of the collective properties of helical geometry of space-time trajectories on $I_U$. For example, as discussed in Section \ref{dirac}, these helical structures can be linked with the spinors which provide the most basic representations of the Maxwell equations \cite{PenroseRindler}.  (It can be noted that the theory of p-adic manifolds as Lie Groups is well established \cite{Schneider}.) Hence, for example, the vacuum fluctuations of quantum field theory can be considered as describing variations in the $k$th iterate symbolic labels associated with a particular $k-1$th iterate trajectory (c.f. Fig \ref{magglass}). As discussed below, relative to the p-adic metric, the distance between individual $k$th iterates is a factor $p$ smaller than the distance between the `gravitationally clumped' regions $a$ and $\cancel a$ discussed above. This suggests that these vacuum fluctuations should be considered as gravitationally indistinct. This has important implications, discussed below. 

The fundamentally different representations of gauge fields and gravity on $I_U$ suggests that it is misguided to imagine that the synthesis of quantum and gravitational physics can be brought about by applying quantum field theoretic ans\"{a}tze to a gravitational Lagrangian. Indeed one can go further: Invariant Set Theory predicts the non-existence of the graviton and hence one will not be discovered in a particle accelerator or elsewhere (c.f. http://home.cern/about/physics/extra-dimensions-gravitons-and-tiny-black-holes). The existence of the graviton has also been called into question by Dyson \cite{Dyson:2013}.

The existence of the spin-2 graviton is sometimes used to support the concept of supersymmetry - the argument being that it is unlikely that there is a gap in the spin $k/2$ particles from $k=0$ to $k=2$. However, this argument fails if in fact there is no spin 2 particle, as predicted is the case above. Suppose indeed that the highest spin elementary particle is spin 1 and supersymmetry is not part of the laws of physics. How then can one account for the dark universe? Here one can draw on the basic concept of Invariant Set Theory described in (\ref{generalisedequation}) (and used to motivate the de Broglie relationships): that the physics of our space-time $\mathcal M_U$ is determined by the geometry of state space in the neighbourhood $\mathcal N$ of $\mathcal M_U$ on $I_U$. This suggests a  generalisation of the dynamical equations of general relativity from 
\be
\label{gr}
G_{\mu \nu} (\mathcal M_U)= \frac{8 \pi G}{c^4} T_{\mu \nu} (\mathcal M_U)
\ee
to
\be
\label{gr2}
G_{\mu \nu}(\mathcal M_U)= \frac{8 \pi G}{c^4} \int_{\mathcal N (\mathcal M_U)} T_{\mu \nu}(\mathcal M_U') F(\mathcal M_U, \mathcal M_U') \ d\mu
\ee
where $F(\mathcal M_U, \mathcal M_U')$ is some propagator (to be determined) and $d \mu$ a suitably normalised Haar measure in some neighbourhood $\mathcal N (\mathcal M_U)$ on $I_U$. Just as the presence of distant matter (as represented by General Relativity Theory) does not destroy the notion of causal order in Special Relativity Theory, so the generalisation (\ref{gr2}) does not affect the notion of causal order in $\mathcal M_U$. That is to say, (\ref{gr2}) is a locally causal extension of the field equations of General Relativity. 

The generalisation (\ref{gr2}) predicts that the integrated (or `smeared out') effect of energy-momentum in space-times neighbouring $\mathcal M_U$ on $I_U$ would influence the curvature of $\mathcal M_U$. Such an influence could be interpreted (wrongly, according to Invariant Set Theory) as implying the existence of  `dark matter' in $\mathcal M_U$; as such, Invariant Set Theory predicts that dark-matter particles will not be discovered experimentally.  
Extension (\ref{gr2}) also provides a potential explanation for dark energy. That is to say, a  basic aspect of the geometry of $I_U$, as shown in Fig \ref{magglass}, is the exponential divergence of nearby space-times. Such divergence occurs ubiquitously in state-space in the neighbourhood of $\mathcal M_U$. Through (\ref{gr2}), this exponential divergence should leave an imprint on our space-time as a positive cosmological constant. This raises the question as to why, in Invariant Set Theory, vacuum fluctuations do not contribute to a vastly larger value of dark energy than is observed (as they do when computed using standard quantum theory). The answer has already been suggested above: the p-adic metric $g_p(\mathcal M_U, \mathcal M_U')$ on $I_U$ does not vary continuously as $\mathcal M_U \rightarrow \mathcal M_U'$, but rather jumps by factors of $p$. This means that when $\mathcal M_U'$ is sufficiently close to $\mathcal M_U$ it effectively has no distinct role in contributing to the integral on the right hand side of (\ref{gr2}). As mentioned, the space-times $\mathcal M_U'$ associated with what in quantum field theory would be described as vacuum fluctuations in $\mathcal M_U$, are examples where the p-adic distance $g_p(\mathcal M_U, \mathcal M_U')$ is so small that it has no impact in (\ref{gr2}).  Whether these speculations stand up to quantitative analysis remains to be seen - in any case, one should recall that these speculations arose from an attempt to formulate a realistic theory of quantum physics, and not to solve the dark universe problem \emph{per se}. 

Invariant Set Theory also provides a novel explanation of the fate of information in black holes. For $I_U$ to be a compact set in $U$'s state space, there must be corresponding regions of state space where trajectories are converging. It seems plausible to suppose that this occurs on Planck scales (an aspect of Invariant Set Theory not discussed in this paper). Consistent with this, Penrose \cite{Penrose:2010} argues that state-space trajectory convergence is generic at final space-time singularities. However, such convergence need not imply loss of information (as Penrose assumes to be the case). The concept of information is closely linked to entropy and conventionally requires one to consider a coarse-graining of state-space volumes. In this way, one can equate information loss to state-space volume shrinkage associated with trajectory convergence. However, the volume of $I_U$ is strictly zero, and hence volumes cannot shrink on $I_U$. That is to say, the resolution of the black hole suitably information paradox may actually be no different to the resolution of the other quantum paradoxes discussed in this paper. For example, in Section \ref{heisenberguncertainty}, the counterfactual Mach-Zehnder experiments that would, if physically real, contradict Laplacian addition of probability, lie off $I_U$. Hence Laplacian addition of probability is not really violated - even though it may seem to be. Similarly, the counterfactual experiments necessary to derive the Bell inequality also lie off $I_U$ and do not correspond to states of physical reality. Hence, local causality is not really violated, even though it may seem to be. Finally, the counterfactual experiments necessary to infer state space volume shrinkage and hence black-hole information loss, also lie off $I_U$ and do not correspond to states of physical reality. Hence, `information' (suitable defined on $I_U$) is not really lost, even though it may seem to be. Notice that we have managed to retain information without needing a firewall or any such structure for the (locally non-computable) event horizon. 

In a late universe with many black holes forming, this suggests would be strong state-space convergence and hence a reversal of dark energy similar to quintessence fields \cite{Zlatev}. This would lead to a strong reduction in what could be described as total entropy. The convergence of space time trajectories associated with a universe collapsing towards a big crunch would therefore be in a low-entropy state before it reemerged into the next epoch. As such, Invariant Set Theory has no need for an inflationary epoch at the time of the Big Bang.

One of the reasons for seeking a quantum theory of gravity is that it should eliminate precise singularities in $\mathcal M_U$. Here we speculate that the `smearing effect' of the generalisation (\ref{gr2}) could do just this: space-times $\mathcal M_U'$ in the neighbourhood $\mathcal M_U$ where a massive star would otherwise collapse towards a singularity, would be collapsing in subtly different ways and the integrated impact of these alternatives on the space-time curvature of $\mathcal M_U$ could prevent the singularity in $\mathcal M_U$ from forming. Such a speculation, like the others in this Section, must of course be put on a proper quantitative footing. 

\section{Conclusions}
\label{conclusions}

A motivating belief underpinning this work is that quantum mechanics is a fundamentally inaccurate theory with which to describe and hence incorporate the phenomenon of gravity. An alternative geometric $\psi$-epistemic theory of quantum physics - Invariant Set Theory - has been described which, it is claimed, can incorporate Einstein's causal geometric theory of gravity straightforwardly. As discussed, this has a number of potentially observable consequences: gravity as an inherently decoherent phenomenon, the non-existence of the graviton, no need for supersymmetry to explain the dark universe, and no need for an inflationary phase of the universe. 

These predictions run completely counter to those of contemporary physical theory. The reason for this is Invariant Set Theory does not conform to the conventional reductionist approach to theory. Instead, Invariant Set Theory is based on a `top-down' postulate that the universe $U$ is a deterministic dynamical system evolving precisely on a measure-zero fractal subset $I_U$ of $U$'s state space. That is to say, it is proposed that the most primitive expression of the laws of physics describe the global fractal geometry of $I_U$, rather than differential evolution equations in space-time. 

Any $\psi$-epistemic theory of quantum physics, is necessarily constrained by no-go theorems (e.g. \cite{Pusey}), which appear to make it inconsistent with experiment. However, making use of the homeomorphism between $I_U$ and the space of p-adic integers, for large but finite $p$, it is shown that these theorems are nullified by using the p-adic metric (rather than the more intuitive Euclidean metric) in state space to distinguish between physically allowable and physically inconsistent counterfactual perturbations. Importantly, neither local causality nor realism are abandoned in Invariant Set Theory. Fractal geometry has many links with number theory, and elementary number theory plays an important role in Invariant Theory, in explaining the principal features of quantum theory, including, for example, the non-commutativity of quantum observables, the violation of the Bell inequalities, and the relationship between probability and frequency of occurrence in space-time. The problem of synthesising quantum and gravitational physics becomes one of combining the locally Euclidean structure space-time with the locally p-adic structure of state space. A way to do this has been proposed in the paper. 

The complex multi-qubit Hilbert Space and the Schr\"{o}dinger equation (here discussed in relativistic form) emerges from Invariant Set Theory in the singular limit \cite{Berry} where (and only where) the otherwise finite $p$ is set equal to infinity. This notion of a singular limit should not be considered pathological. Indeed, Berry \cite{Berry} notes that singular limits are commonplace in science, and provide insight into how a more general theory (e.g. the Navier-Stokes theory of viscous fluids) can reduce to a less general theory (e.g. the Euler theory of inviscid fluids), and therefore how higher-level phenomena can emerge from lower-level ones. 
In such situations, the approximate theory can describe reality very accurately for many purposes, but fail catastrophically in other situations. Many aspects of fluid turbulence in high Reynolds number flows can be described well by the Euler equations. However, Euler theory fails utterly in describing the phenomenon of heavier-than-air flight! Here it is suggested that whilst quantum theory is an excellent predictor of laboratory experiments, it may fail catastrophically in situations where the effects of gravity are nontrivial. Whilst many of these failures may only be strongly apparent on the cosmological scale, the author looks forward to the execution of laboratory-scale experiments in coming years which can test the potentially decoherent nature of gravity during quantum measurement. 

\section*{Acknowledgement} My thanks to Harvey Brown, Shane Mansfield, Simon Saunders, Kristian Strommen and David Wallace and anonymous reviewers for helpful comments on an earlier version of this paper. 
\bibliography{mybibliography}

\appendix

\section{Multiple Qubits}
\label{MQ}

A general $m$ qubit state can be built from a general $m-1$ qubit state using the following inductive formula:
\begin{eqnarray}
\label{Mqubit2}
&|\psi_{\underbrace{a, b \ldots d}_{m}}(\theta_1, \ldots \theta_{2^m-1}; \phi_1, \ldots \phi_{2^m-1})\rangle=\nonumber\\
&\cos \frac{\theta_1}{2}|a\rangle
\times |\psi_{\underbrace{b, c \ldots d}_{m-1}}(\theta_2, \ldots \theta_{2^{m-1}}; \phi_2 \ldots \phi_{2^{m-1}}) \nonumber \\
&+\sin \frac{\theta_1}{2} e^{i\phi_1}|\cancel{a}\rangle \times |\psi_{\underbrace{b, c \ldots d}_{m-1}}(\theta_{{2^{m-1}+1}} \ldots \theta_{2^m-1};\phi_{{2^{m-1}+1}} \ldots \phi_{2^m-1})\rangle
\end{eqnarray}
The correspondence with bit strings is similarly defined inductively. Let
 \begin{eqnarray}
 \label{NM1bitstrings}
&|\psi_{b,c \ldots d}(\theta_2, \ldots \theta_{2^{m-1}}; \phi_2 \ldots \phi_{2^{m-1}}) \rangle \nonumber \\
\\
 &\mapsfrom
\begin{dcases}
S_{b}(\theta_2, \phi_2)=\{b'_1, b'_2, b'_3, \ldots b'_{2^N}\} \\
S_{c}(\theta_3, \phi_3)=\{c'_1, c'_2, c'_3, \ldots c'_{2^N}\}\\
\ldots \\
S_{d}(d; \theta_{2^{m-1}}, \phi_{2^{m-1}})=\{d'_1, d'_2, d'_3, \ldots d'_{2^N}\}
\end{dcases}
\end{eqnarray}
and 
\begin{eqnarray}
\label{NM1bitstrings}
&|\psi_{b,c \ldots d}(\theta_{2^{m-1}+1}, \ldots \theta_{2^m-1}; \phi_{2^{m-1}+1} \ldots \phi_{2^m-1}) \rangle \nonumber \\
\\
&\mapsfrom
\begin{dcases}
S_{b}(\theta_{2^{m-1}+1}, \phi_{2^{m-1}+1})=\{b''_1, b''_2, b''_3, \ldots b'_{2^N}\} \\
S_{c}(\theta_{2^{m-1}+2}, \phi_{2^{m-1}+2})=\{c''_1, c''_2, c''_3, \ldots c''_{2^N}\}\\
\ldots \\
S_{d}(\theta_{2^m-1}, \phi_{2^m-1})=\{d''_1, d''_2, d''_3, \ldots d''_{2^N}\}
\end{dcases}
\end{eqnarray}
define a pair of $m-1$ qubit correspondences (where $b'_i$, $b''_i \in \{b, \cancel b\}$ etc), then with 
\be
\cos\frac{\theta_1}{2}\;|a\rangle + \sin\frac{\theta_1}{2} e^{i \phi_1}\;|\cancel{a}\rangle \mapsfrom  \{a_1, a_2, a_3 \ldots a_{2^N}\},
\ee
an independent bit string with $a_i \in \{a, \cancel a\}$, the $m$ qubit correspondence can be written as
\be
 \label{Nbitstrings}
|\psi_{a,b \ldots d}(\theta_1, \ldots \theta_{2^m-1}; \phi_1 \ldots \phi_{2^m-1}) \rangle \mapsfrom
\begin{dcases}
\{a_1, a_2, a_3, \ldots a_{2^N}\} \\
\{b_1, b_2, b_3, \ldots b_{2^N}\} \\
\{c_1, c_2, c_3, \ldots c_{2^N}\}\\
\ldots \\
\{d_1, d_2, d_3, \ldots d_{2^N}\}
\end{dcases}
\ee
where
\begin{align}
b_i &= b'_i \;\; c_i=c'_i \;\; \ldots d_i=d'_i&\mathrm{if}& \;\; a_i=a \nonumber \\
b_i &= b''_i \;\; c_i=c''_i \;\; \ldots d_i=d''_i&\mathrm{if}& \;\;a_i = \cancel{a}
\end{align}
and $b_i \in \{b, \cancel b\}, c_i \in \{c, \cancel c\}$ etc. This reduces to (\ref{condition}) when $m=2$. 
\bigskip

\end{document}